\documentclass{cimento}
\usepackage{graphicx}        
\usepackage{color}           
\usepackage{url}             

\title{Solar Limb Darkening Function from Baily's Beads Observations}

\author{A.~Raponi\from{ins:s}\ETC,
        C.~Sigismondi\from{ins:s}\from{ins:i}\from{ins:e}}

   \instlist{\inst{ins:s} Sapienza University of Rome, P.le Aldo Moro 5 00185, Roma (Italy)  
             \inst{ins:i} ICRA, International Center for Relativistic Astrophysics, Roma (Italy) {\rm sigismondi@icra.it}
             \inst{ins:e} University of Nice-Sophia Antipolis (France); IRSOL, Istituto Ricerche Solari di Locarno (Switzerland) and GPA Observatorio Nacional (Rio de Janeiro, Brasil)}
\PACSes{\PACSit{95.10.Gi }{ Eclipses}{ 96.60.-j}{ - Solar physics.  }{96.60.Bn  }{ - Solar diameter.}}

\begin{document}

\maketitle

\begin{abstract}
We introduce a method to measure with high resolution the solar diameter from the ground, through the eclipse observations by reconsidering the definition of the solar edge. The outer part of the Limb Darkening Function (LDF) is recovered using the luminosity evolution of a Baily's Bead and the profile of the lunar limb available from the Kaguya satellite. 
The method proposed is applied for the videos of the eclipse in January, 15, 2010 recorded by Richard Nugent in Uganda and Andreas Tegtmeier in India. The result shows light from solar limb detected at least 0.65 arcsec beyond the LDF inflection point, and this fact may suggest to reconsider the evaluations of the historical eclipses made with naked eye. 
\end{abstract}

\section{The method of eclipses}

With the eclipse observations we are able to bypass some problems that affect the measurement of solar diameter. The atmospheric and instrumental effects that distort the shape of the limb \cite{Djafer} are overcome by the fact that the scattering of the Sun's light is greatly reduced by the occultation of the Moon, therefore there are much less photons from the photosphere to be poured, by the PSF effect, in the outer region. 

\subsection{Comparison between observation and ephemeris} 
The method exploits the observation of the beads of light that appear or disappear from the bottom of a lunar valley when the solar limb is almost tangent to the lunar limb. 

The timing of their appearing or disappearing and their positions are measured. These instants when the photosphere disappears or emerges behind the valleys of the lunar limb, are determined solely by the positions of the Sun and the Moon with respect to the observer; their angular size depends on the lunar marginal profile while the effects of the atmospheric seeing on these phenomena are negligible.

The International Occultation Timing Association (IOTA) is currently engaged to observe the eclipses with the aim of measuring the solar diameter. This is facilitated by the development of the software Occult 4 by David Herald\footnote[1]{\url{www.lunar-occultations.com/iota/occult4.htm}}.

The technique consists to look at the time of appearance of the beads and to compare it with the calculated positions by the ephemeris using the software Occult 4. The simulated Sun by Occult 4 has the standard radius: 959.63 arcsec at 1AU \cite{Auwers}. The difference between the simulations and the observations is a measure of the radius correction with respect to the standard radius ($\Delta$R).

In the lunar polar regions, due to the geometry of the eclipse, the beads last for more time \cite{Sigi2009}. Observing in the lunar polar regions we are also able to avoid the measurement in the solar active regions that could affect the measures \cite{Thuillier}. In fact the solar active regions do not appear in solar latitude higher than $\sim 40^\circ$. The maximum offset between the lunar and the solar pole is $\sim 9^\circ$. 

However this approach conceives the bead as an on-off signal. In this way one assumes the Limb Darkening Function as a step function, but it is actually not. Different compositions of optical instruments (telescope + filter + detector) could have different sensitivities and different signal/noise ratios, recording the first signal of the bead in correspondence of different points along the luminosity profile. This leads to different values of the perceived $\Delta$R i.e. the correction to the standard angular solar radius at 1 AU (see Fig. 1).
An improvement of this approach has to take into account the whole shape of the Limb Darkening Function and thus the actual position of the inflection point.

\begin{figure}
\centerline{\includegraphics[width=0.5\textwidth,clip=]{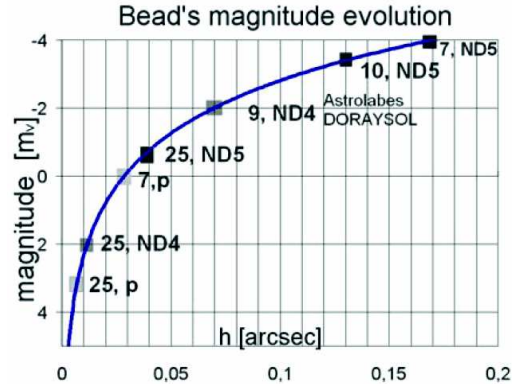}}
\caption{Bead´s magnitude evolution: the height of the solar limb above the valley is on the abscissa. The various square dots represent different type of telescopes. [25, p] means 25 cm opening with projection of the image. [25, ND4] the 25 cm telescope with a filter of Neutral Density of transmittance 1/10000; ND5 stands for 1/100000, and so on.}
\label{Fig. 1}
\end{figure}

\subsection{Numerical calculation}

The shape of the light curve of the bead is determined by the shape of the LDF (not affected by seeing) and the shape of the lunar valley that generates the bead. 
Calling w(x) the width of the lunar valley (i.e. the length of the solar edge visible from the valley in function of the height x from the bottom of the valley), and B(x) the surface brightness profile (i.e. the LDF), one could see the light curve L(y) as a convolution of B(x) and w(x), being $\mid y\mid$ the distance between the botton of the lunar valley and the standard solar edge, setting to 0 the position of the standard edge.

\noindent $L(y)=\intop B(x)\, w(y-x)\, dx$

\noindent The discrete convolution is:

\noindent $L(m)=\sum B(n)\, w(m-n) h$

\noindent where n, m are the index of the discrete layers corresponding to x, y coordinate and h the layers thickness.

Thus the profile of the LDF is discretized in order to obtain the solar layers B(n) of equal thickness and concentric to the center of the Sun. In the short space of a lunar valley these layers are roughly parallel and straight. 

The lunar valley is also divided in layers of equal angular thickness. During a bead event every lunar layer is filled by one solar layer every given interval of time. Step by step during an emerging bead event, a deeper layer of the solar atmosphere enters into the profile drawn by the lunar valleys (see Fig. 2), and each layer casts light through the same geometrical area of the previous one. 

\begin{figure}
\centerline{\includegraphics[width=0.6\textwidth,clip=]{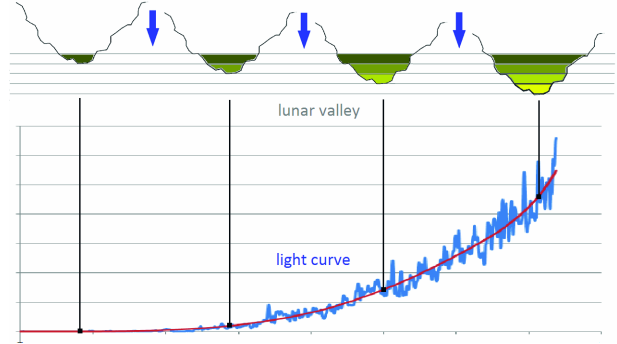}}
\caption{Every step in the geometry of the solar-lunar layers (up) corresponds to a given instant in the light curve (down). The value of the light curve is the contribute of all the layers.}
\label{Fig. 2}
\end{figure}

Being: $A_{1},\, A_{2}..\: A_{n}$ the area of lunar layers, from the bottom of the valley going outward; $B_{1},\, B_{2}..\: B_{n}$ the surface brightness of the solar layers (our goal) from the outer (dimmer) going inward; $L_{1},\, L_{2}..\: L_{n}$ the value of the observed light curve from the first signal to the saturation of the detector or to the replenishment of the lunar valley. One has: 

\noindent $L_{m} = B_{1}A_{m} + B_{2}A_{m-1} + .. + B_{m}A_{1} = \sum_{n=1}^m B_{n}\, A_{m-n+1}$

To derive the LDF profile:

\noindent $B_{1}=L_{1}/A_{1}$

\noindent $B_{2}=\left[L_{2}-(B_{1}\cdot A_{2})\right]/A_{1}$

\noindent $B_{3}=\left[L_{3}-\left(B_{1}\cdot A_{3}+B_{2}\cdot A_{2}\right)\right]/A_{1}$

\noindent $B_{4}=\left[L_{4}-\left(B_{1}\cdot A_{4}+B_{2}\cdot A_{3}+B_{3}\cdot A_{2}\right)\right]/A_{1}$ 

\noindent and so on. 

\noindent The situation described above relates to an emerging bead. The same process can run for a disappearing bead, simply plotting the light curve back in time.

\section{An application of the method of eclipses}
We studied the videos of the annular eclipse in January 15, 2010 realized by Richard Nugent in Uganda and Andreas Tegtmeier in India. 

The equipment of Nugent was: CCD camera Watec 902H Ultimate\footnote[2]{\url{http://www.aegis-elec.com/products/watec-902H_spec_eng.pdf}}; Matsukov telescope (90/1300 mm); panchromatic filter Thousand Oaks, ND5. 

The equipment of Tegtmeier was: CCD camera Watec 120N\footnote[3]{\url{http://www.aegis-elec.com/products/watec-100N_spec_eng.pdf}}; Matsukov telescope (100/1000 mm); Filter: I = IOTA/ES green glass based neutral ND4. 

\subsection{Analysis}
Two beads located at Axis Angle\footnote[4]{the angle around the limb of the Moon, measured Eastward from the Moon's North pole} (AA) $=171^\circ and 177^\circ$ are analyzed for both the videos. 

The lunar valley analysis is performed with the new lunar profile obtained by the laser altimeter (LALT) onboard the Japanese lunar explorer Kaguya\footnote[5]{\url{http://wms.selene.jaxa.jp/selene_viewer/index_e.html}}. The radial topographic error is estimated to be ± 4.1 m corresponding to 2.1-2.3 mas at the distance of the Earth. The choice of the thickness of the layers has to be optimal: large enough to reduce $B_n$ uncertainties, but small enough to have a good resolution of the LDF. We chose $h=30$ mas for the lunar valley at AA = 177$^\circ$ and $h=73$ mas for the lunar valley at AA = 171$^\circ$.

The motion of the Sun in the lunar valley is simulated with the Occult 4 software. 

A program in Fortran90 is performed to calculate the LDF points ($B_n$), taking into account the light curves of the beads, the lunar profiles and the respective uncertainties.


\subsection{Results}
The resulting points show that the inflection point is clearly between the two profile obtained for each of the two beads. The saturation of the CCD pixels avoided to measure the luminosity function more inward for Nugent's video, while the low sensitivity avoided to measure the luminosity function more outward for Tegtmeier's video. Therefore from these points it is impossible to infer an exact location of the inflection point. But it is possible to deduce an upper and lower limit corresponding to the points that better constrain it: -0.190 arcsec $< \Delta R <$ +0.050 arcsec. 

The analysis of these beads with the "classical approach" would lead to a difference greater than one arcsecond for $\Delta$R between the 2 videos. The improvement introduced by this approach is then shown.

\begin{figure}
\centerline{\includegraphics[width=0.5\textwidth,clip=]{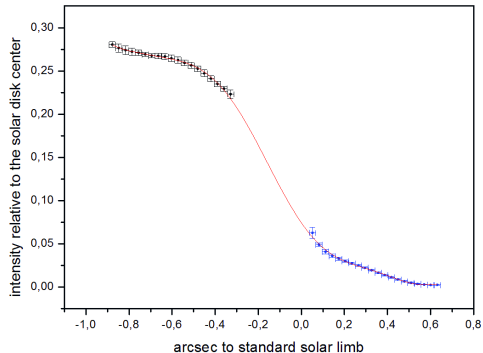}} 
\caption{The luminosity profiles obtained for the bead at AA = 177$^\circ$ are plotted and put together. The inner and brighter part is obtained from Tegtmeier's video; the outer and weaker part is obtained from Nugent's video. The luminosity profile is normalized to the center of the solar disk according to Rogerson \protect\cite{Rogerson} for the inner part, and in arbitrary way for the outer part. The zero of the abscissa is the position of the standard solar limb with a radius of 959.63 arcsec at 1 AU. The error bars on y axis are the 90\% confidence level. The error bars on x axis are the thickness (h) of the lunar layers. The solid line is an interpolation between the profiles and gives a possible scenario on the position of the inflection point.}
\label{Fig. 4}
\end{figure}

\section{Historical eclipses}

Even if observed with the naked eyes, historical eclipses are of great interest for studying the variations of the solar diameter over periods longer than a solar cycle. 

As an example in April 9, 1567 in Rome, Chrispopher Clavius observed an annular eclipse \cite{Clavius}. The solar limb should have been at least $\Delta R > +2.5$ arcsec in order to be higher than the mountains of the Moon.
Eddy \etal  \cite{Eddy} deduced from this interpretation a secular shrinking of the Sun from the 16th century to the present.
In general the evaluations of historical eclipses is made without any reference to the inflection point, and considering the LDF as a Heaviside profile. This can enlarge the measured $\Delta$R. It is the main concern with naked eye observations.  


\section{Conclusions}

We obtained a detailed profile of the outer part of the LDF, demonstrating the functionality of the method. 
Although it was impossible to observe the inflection point, a consideration is coming out: if the LDF is considered as a step function, like up to now in the eclipse analyses, the solar diameter can be overstimated.
And this effect can apply to historical eclipses, of which we reconsidered the Clavius' one (1567).
Moreover the LDF profile, around the inflection point position, is an important source of information for the solar atmosphere \cite{Thuillier}.
Improvements to obtain a detailed profile comparable to the space astrometry are possible:
\begin{itemize}
\item{An increased dynamic range of the CCD detectors from 8 bits to 12 bits is recommended in order to avoid CCD saturation before reaching the inflection point.} 
\item{The profile is dependent on the wavelength \cite{Djafer,Thuillier}. One must then observe at specific photometric bands to be able to compare the results with other measures.} 
\end{itemize}


\bibliography{LDF}
   
\bibliographystyle{varenna}

\end{document}